\documentclass[final,authoryear,5p,times,twocolumn]{elsarticle}
\usepackage{epsfig}
\usepackage{amssymb}
\journal{Journal of Theoretical Biology}

\begin{document}

\begin{frontmatter}

\title{Coexistence of fraternity and egoism for spatial social dilemmas}

\author[mfa]{Gy\"orgy Szab\'o}
\author[mfa]{Attila Szolnoki}
\author[elte]{Lilla Czak\'o}

\address[mfa]{Institute of Technical Physics and Materials Science, Research Centre for Natural Sciences, Hungarian Academy of Sciences, P.O. Box 49, H-1525 Budapest, Hungary} 
\address[elte]{Roland E{\"o}tv{\"o}s University, Institute of Physics, P{\'a}zm{\'a}ny P. s{\'e}t{\'a}ny 1/A, H-1117 Budapest, Hungary}

\begin{abstract}
We have studied an evolutionary game with spatially arranged players who can choose one of the two strategies (named cooperation and defection for social dilemmas) when playing with their neighbors. In addition to the application of the usual strategies in the present model the players are also characterized by one of the two extreme personal features representing the egoist or fraternal behavior. During the evolution each player can modify both her own strategy and/or personal feature via a myopic update process in order to improve her utility. The results of numerical simulations and stability analysis are summarized in phase diagrams representing a wide scale of spatially ordered distribution of strategies and personal features when varying the payoff parameters. In most of the cases only two of the four possible options prevail and may form sublattice ordered spatial structure. The evolutionary advantage of the fraternal attitude is demonstrated within a large range of payoff parameters including the region of prisoner's dilemma where egoist defectors and fraternal cooperators form a role-separating chessboard-like pattern.
\end{abstract}


\end{frontmatter}

\section{Introduction}
\label{intro}

Multi-agent game theoretical models give us a general mathematical tool to describe real-life situations in human societies and to study biological systems when varying the interactions, evolutionary rules, and connectivity structure among the players \citep{maynard_82,nowak_06,sigmund_10,pacheco_jtb08}. In many cases the interactions are approximated by the sum of pair interactions between neighboring (equivalent) players distributed on the sites of a lattice or graph (for a survey see \citep{nowak_ijbc93,szabo_pr07,perc_bs10}). 
The simplest spatial versions of two-strategy games have demonstrated new outcomes of evolutionary process, which are missing if well-mixed players are postulated.

To give an example, the most exhaustively studied symmetric two-person two-strategy game is the so-called Prisoner's Dilemma (PD) game where the equivalent players can choose cooperation or defection. For mutual cooperation (defection) both players receive a payoff $R$ ($P$) while for their opposite decisions the cooperator (defector) gains $S$ ($T$). For the PD game the payoffs satisfy the conditions: $S<P<R<T$, that enforces both selfish players to choose defection (representing the state called the ''tragedy of the commons'' \citep{hardin_g_s68}) meanwhile the mutual cooperation would be more beneficial for the players. 
Being trapped in the state of mutual defection is in stark contrast to our everyday experience of high level of cooperation. To resolve this discrepancy several cooperation supporting conditions and
mechanisms were identified
\citep{nowak_s06,pacheco_ploscb06,fu_pre09,poncela_njp09,pacheco_jtb06,tomassini_bs10,gomez-gardenes_jtb08,fort_epl08,perc_pre11,vukov_pone11,fu_srep12,pinheiro_pone12}. 
Alternative ways to explain the emergence of cooperation are originated from the observation that humans follow more complex behavior that cannot be well described by simple unconditional cooperator and defector acts. Human experiments \citep{fehr_eer02,camerer_03,nowak_06,sigmund_10,traulsen_pnas10} 
highlighted that individuals possess
different personal features or emotions \citep{szolnoki_epl11}, e.g., selfish \citep{neumann_44}, altruistic \citep{sigmund_sa02}, fraternal \citep{scheuring_jtb10,szabo_jtb12}, punishing \citep{clutton_brock_n95,fehr_n02,kurzban_pnas05}, reciprocative \citep{berg_geb95}, envy \citep{garay_bs11,szolnoki_epl11}, just to name a few examples. Following this avenue, now we introduce a spatial
model where players are not limited to the use of the pure cooperator and defector strategies but they are also motivated by an additional personal feature characterizing their egoist or fraternal attitude.  Accordingly, the present work 
generalizes and extends previous specific efforts about the consequences of collective decisions \citep{szabo_pre10} and other-regarding preferences for a uniform level of fraternal behavior \citep{szabo_jtb12}. Here it is worth mentioning that the
fraternal behavior can prevent the society from falling into the "tragedy of commons" state. Consequently, the advantage of the fraternal behavior can be interpreted as an evolutionary driving force supporting societies to maintain/develop the altruistic personal features. Studying the present model we wish to explore the consequences of the spatial competition (evolutionary process) between the above described strategy profiles. It is emphasized, furthermore, that for the quantum games \citep{abal_pa08,li_q_pa12} the players exhibit a behavior similar to those played by fraternal players.

For the sake of comparison the present analysis is also performed for all $2 \times 2$ social dilemmas games [including PD, Hawk-Dove (HD) and Stag-Hunt (SH) games] when varying the values of $T$ and $S$ (for $R=1$ and $P=0$ without loss of generality). Finally we mention that the present four-strategy model is analogous to those cases when the spatial social dilemmas are studied by considering voluntary participation \citep{szabo_pre02b}, punishments \citep{rand_jtb09,sekiguchi_jtb09,helbing_ploscb10}, and the use of sophisticated strategies like Tit-for-tat \citep{nowak_n92a} or others \citep{ohtsuki_jtb04,ohtsuki_jtb06b,rand_jtb09}.

Due to the biological motivations \citep{maynard_82} in the early evolutionary games the time-dependence of the strategy distribution is controlled by the imitation of a better performing neighbor. In human societies, however, we can assume more intelligent players who are capable to evaluate their fictitious payoff variation when modifying
strategy \citep{sysiaho_epjb05,szabo_pr07}. The corresponding so-called
myopic evolutionary rule is analogous to the Glauber dynamics \citep{glauber_jmp63} used frequently in the investigation of stationary states and dynamical processes for the kinetic Ising model \citep{binder_88}. In biological systems the latter mechanism can be interpreted as the survival of possible mutants with a probability increasing with the current fitness. Contrary to the imitation of a neighbor, the mentioned myopic dynamical rule permits the formation of sublattice ordered distribution of strategies (and/or personalities) resembling the anti-ferromagnetic structure in the Ising systems. For the case of spatial PD the chessboard like arrangement of cooperators and defectors is favored if $T+S>2R$. This latter criterion coincides with those one when the players have the highest average income in the repeated two-person PD game if they alternate cooperation and defection in opposite phase.

It will be demonstrated that the present model exhibits different disordered and sublattice ordered spatial arrangements as well as phase transitions when varying the payoff parameters for several fixed levels of noise. It means that in contrary to preliminary/naive expectation the fraternal players may survive in the presence of egoist competitors.

The rest of this paper is organized as follows. First, we describe our four-strategy lattice model. The results of Monte Carlo (MC) simulations at a fixed noise level are detailed in Sec.~\ref{mc}
while phase diagram in the low noise limit are discussed in Sec.~\ref{sec:pdlnl}. This diagram can be obtained by means of 
stability analysis of the possible two-strategy phases against the point defects. The essence of this method and an analytical estimation for the direction of interfacial invasion between the mentioned phases are briefly described in Sec.~\ref{stabanal}. Finally, we summarize our main finding and discuss their implications.

\section{The model}
\label{model}

In the present model the players are located on the sites of a square lattice with $L \times L$ sites. The undesired effects of boundaries are eliminated by using periodic boundary conditions in the simulations. At each site $x$ four types of players are distinguished, namely, $s_x=D_e$, $C_e$, $D_f$, and $C_f$. In our notation $D_e$ and $C_e$ refer to egoist defector and cooperator while $D_f$ and $C_f$ denote fraternal defector and cooperator for the PD games. For other types of games ({\it e.g.}, HD or SH) we will use the above-mentioned abbreviations of types that we call strategies henceforth. In the mathematical formulation of utilities these strategies are denoted by the following unit vectors:    
\begin{equation}
\label{eq:state4}
D_e=\left( \matrix{1\cr 0\cr 0\cr 0} \right), \; 
C_e=\left( \matrix{0\cr 1\cr 0\cr 0} \right), \;
D_f=\left( \matrix{0\cr 0\cr 1\cr 0} \right), \;
C_f=\left( \matrix{0\cr 0\cr 0\cr 1} \right). 
\end{equation}

The utility $U({\bf s}_x)$ of the player $x$ (with a strategy ${\bf s}_x$) comes from games with her four nearest neighbors and can be expressed by the following sum of matrix products:
\begin{equation}\label{ui}
U({\bf s}_x) = \sum_{\delta} {\bf s}_x \cdot {\bf A} {\bf s}_{x+\delta}.
\end{equation}
Here the summation runs over the four nearest neighboring sites of $x$ and the payoff matrix ${\bf A}$ is given as
\begin{equation}
{\bf A}=\left(\matrix{ 0      & T      & 0      & T      \cr
                       S      & 1      & S      & 1      \cr
                       0      & \sigma & 0      & \sigma \cr
                       \sigma & 1      & \sigma & 1  }\right)
\label{eq:pom4s}
\end{equation}
where $\sigma = (T+S)/2$. The matrix elements in the upper-left $2 \times 2$ block of the whole payoff matrix define the payoffs between egoist players ($D_e$ and $C_e$). The present notation is adopted from the literature of social dilemmas where $T$ refers to "temptation to choose defection", $S$ is abbreviation of "sucker's payoff", the "punishment for mutual defection" is chosen to be zero ({\it i.e.}, $P=0$), and the "reward for mutual cooperation" is set $R=1$ for a suitable unit. In contrary to the egoist individuals the fraternal players revalue their payoffs by assuming equal sharing of the common income. In other words, the fraternal players wish to maximize their common income therefore their utility is expressed by a partnership game \citep{hofbauer_98} represented by the lower-right $2 \times 2$ block of the payoff matrix (\ref{eq:pom4s}). It is emphasized that the utility of a given player is based on her own character and is independent of the personal feature (egoist or fraternal) of the co-players. Notice, furthermore, that for the present symmetric game the pair of egoist and fraternal players have the same utility (1 or 0) if both follow the same strategy (cooperation or defection). 

The main advantage of the present approach is that we have only two payoff parameters ($T$ and $S$) when studying the competition between the egoist and fraternal players. Further advantage of this parametrization is that the spatial evolutionary game with only egoist (or fraternal) players were already studied for myopic strategy update at arbitrary payoffs \citep{szabo_pre10,szabo_jtb12} and the results will serve as references for later comparisons. Accordingly, in each elementary step we choose a player $x$ at random and her strategy is changed from ${\bf s}_x$ to ${\bf s}_x^{\prime}$ (chosen at random, too) with a probability  
\begin{equation}
W({\bf s}_x \to {\bf s}_x^{\prime}) =\frac{1}{1+\exp[(U({\bf s}_x)-U({\bf s}_x^{\prime}))/K]} \;,
\label{eq:dyn}
\end{equation}
where $K$ describes the noise amplitude disturbing the players to achieve their optimum utility. 
It is worth noting that not only the ``best myopic response'' is allowed according to the suggested strategy update protocol. Less favorable changes, with smaller probability, are also possible. This freedom of microscopic dynamics makes possible to avoid frozen states reported by \cite{sysiaho_epjb05}.
The iteration of these elementary steps drives the system from any initial state to a final stationary one that is characterized by its own spatial structure and by the strategy frequencies. 

To characterize the wide scale of ordered states we introduced four sublattices (with representative sites forming a $2 \times 2$ block) as it is used in the investigation of several Ising models previously \citep{binder_prb80}. In the stationary state the average strategy frequencies are determined in each sublattices. Both the quantitative analysis and the visualization of the spatial strategy distributions have justified, however, that the behavior of the present system can be well described by a simpler, two-sublattice formalism we use henceforth. More precisely, the square lattice is divided into two sublattices (resembling the chessboard like distinction of sites), $\alpha = 1$ and $2$, and the spatial distribution is described by the strategy frequencies $\rho_{\alpha ,s}$ ($s=D_e,C_e,D_f,$ and $C_f$) in both sublattices. Evidently, $\sum_s \rho_{\alpha, s}=1$ for $\alpha=1,2$. For later convenience in several plots the emerging ordered phase is also characterized by the sum of the total frequencies of two strategies, for example, $\nu(C_e+C_f)=\sum_{\alpha} (\rho_{\alpha, Ce}+\rho_{\alpha,Cf})/2$ denotes the average frequency of $C_e$ and $C_f$ strategies in the whole system.
 
The quantitative analysis has justified that all the possible two-strategy subsystems will play relevant role in the stationary behavior of this system. Two possible subsystems, when only egoist (${\bf s}_x=D_e, C_e$) or fraternal (${\bf s}_x=D_f, C_f$) strategies are permitted, have already been mentioned above \citep{szabo_jtb12}. 
In addition we have two trivial cases, when only cooperative (more precisely ${\bf s}_x=C_e, C_f$) or defective (${\bf s}_x=D_e, D_f$) strategies are permitted in the system. In both cases the players get the same utility independently of the local arrangement of personal features. Consequently, the "cooperative" (or "defective") strategies can coexist in a random, time-dependent structure resembling the high noise limit ($K \to \infty$) for any pair of strategies. For the other two possibilities (${\bf s}_x=D_e, C_f$ or ${\bf s}_x=C_e, D_f$) the corresponding subsystem can be mapped into the case of egoist players with a suitable transformation of payoff parameters.

\section{Monte Carlo results}
\label{mc}
 
The present model is systematically analyzed by MC simulations when varying the payoff parameters ($T$ and $S$) for a fixed noise level, $K=0.25$. During the simulations we have determined the strategy frequencies in both sublattices ($\rho_{\alpha,s}$) and the average payoff per capita in the final stationary state. The system size is varied from $L=400$ to $2000$ in order to have sufficiently accurate statistical data (the statistical errors are comparable to the line thickness). The larger system sizes are used when approaching the critical transitions belonging to the Ising universality class \citep{szabo_pre10}. Simultaneously, the undesired effects of the divergency of relaxation time and of the fluctuation (multiplied by the system size) at the critical points \citep{stanley_71} are reduced by the increase of sampling and thermalization time from $t_s=t_{th}=10^4$ to $10^6$ MCS (1 MCS = $L \times L$ elementary steps described above) when approaching the transition points. As a result, our numerical data exhibit smooth variation even in the vicinity of the critical points. 

If the system is started from a random initial state, then one can observe growing domains yielding the above mentioned sublattice ordered arrangement of strategies. The general features of this phenomenon is already well described in the literature of ordering process (for a survey see \citep{bray_ap94}). As the typical size $l$ of domains grows algebraically with the time (more precisely, $l \propto t^{1/2}$) therefore we need extremely long relaxation time to approach the final (mono-domain) ordered structure for large sizes. The latter difficulty is avoided by starting the consecutive simulations from a prepared, partially ordered initial states whose symmetry agrees with those of the final stationary state. 

\begin{figure}[ht]
\centerline{\epsfig{file=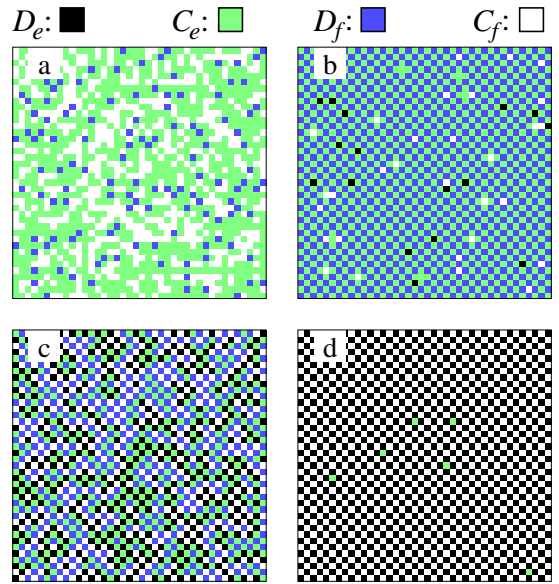,width=7.5cm}}
\caption{Typical ($40 \times 40$) snapshots of strategy distributions in the stationary states for $S=1.5$ and $K=0.25$ if  $T=0.3$ (a), $1.0$ (b), $1.5$ (c), and $2.2$ (d). The strategy colors are indicated at the top.}
\label{config4}
\end{figure}

First we illustrate the variation of strategy distributions by four snapshots obtained for different values of $T$ at fixed values of $S$ and noise $K$. The first snapshot (at $T=0.3$, $S=1.5$, and $K=0.25$) of Fig.~\ref{config4} shows a random distribution of $C_e$ and $C_f$ strategies decorated rarely with the presence of $D_f$ strategy. 
For these payoff values $\rho_{Df} \to 0$ in the zero noise limit. In snapshot b of Fig.~\ref{config4} ($T=1$) one can observe a sublattice ordered arrangement of $C_e$ and $D_f$ strategies with a low frequency of point defects. Surprisingly, for a significantly higher value of $T$ (see snapshot d) the simulations have indicated another sublattice ordered arrangement composed of $D_e$ and $C_f$ strategies. The transition between the latter two patterns is continuous as represented by snapshot c where locally both types of the ordered structures are recognizable. It is more interesting, however, that within the intermediate region $C_f$ strategy is replaced for $C_e$ within one of the sublattices and simultaneously the players with strategy $D_f$ (in the complementary sublattice) modify their strategy to $D_e$. In other words, a sublattice ordered structure is present permanently meanwhile the strategy composition varies smoothly in both sublattices. 

\begin{figure}[ht]
\centerline{\epsfig{file=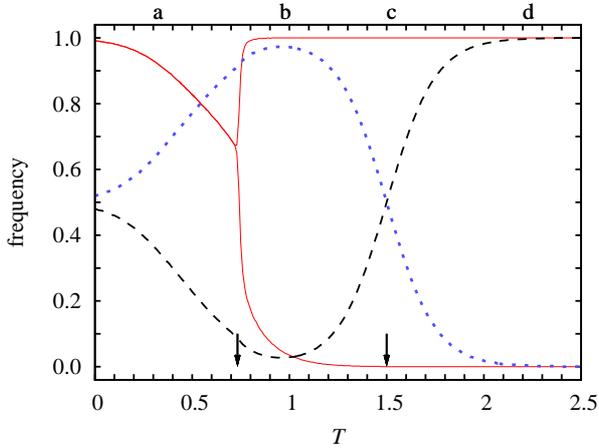,width=8.0cm}}
\caption{Frequency of $C_e$ and  $C_f$ strategies in the sublattices {\it vs.} $T$ are denoted by solid (red) lines for $S=1.5$ and $K=0.25$. The dotted (blue) and dashed (green) lines indicate the quantities $\nu(D_e+C_f)$ and $\nu(D_f+C_e)$ and the arrows show the transition points as explained in the text.}
\label{phtr15}
\end{figure}

The above mentioned process is quantified by an appropriate combination of strategy frequencies (e.g., $\nu(D_e+C_f)$) as illustrated in Fig.~\ref{phtr15} when increasing the value of $T$. At the top of this figure the positions of the labels a, b, c, and d indicate the value of $T$ for which the strategy distributions are shown in the corresponding snapshots of Fig.~\ref{config4}. The consecutive states and transitions are demonstrated by the sum of the frequency of $C_e$ and $C_f$ strategies in both sublattices ($\rho_{\alpha,Ce}+\rho_{\alpha,Cf}$) as a function of $T$ for a fixed $S$ and noise level. 
These curves illustrate clearly the sublattice ordering in the arrangement of strategies if $T$ exceeds a threshold value $T_{c}$ (indicated by the left-hand vertical arrow in Fig.~\ref{phtr15}) dependent on $K$ and $S$. Notice that these quantities do not indicate the transition that becomes visible when plotting the total frequency of $D_e$ and $C_f$ [denoted as $\nu(D_e+C_f)$] strategies as well as $\nu(D_f+C_e)$. The quantities $\nu(D_e+C_f)$ and $\nu(D_f+C_e)$ approach 1 if the two given strategies (namely a cooperative and a defective strategy) form a chessboard like pattern. The right-hand arrow in Fig.~\ref{phtr15} shows the value of $T$ where $\nu(D_f+C_e)$ becomes larger than $\nu(D_e+C_f)$. Within the coexistence region (at a finite $K$ value) the cooperative ($C_e$ and $C_f$) as well as the defective ($D_e$ and $D_f$) strategies are located within the same sublattice.  

\begin{figure}[ht]
\centerline{\epsfig{file=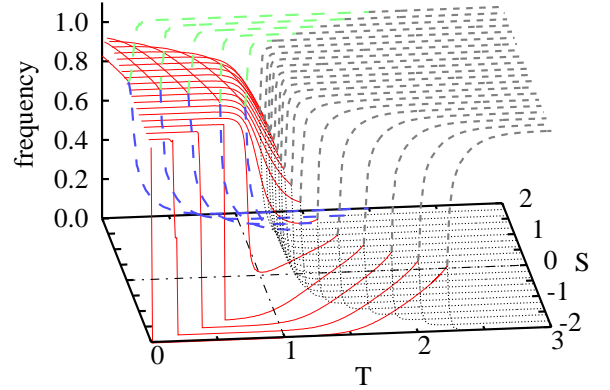,width=7.8cm}}
\caption{Frequency of cooperation (or portion of $C_e$ and $C_f$ strategies) in the sublattices when varying $T$ and $S$ for $K=0.25$. The solid (red) lines refer to the absence sublattice ordering, the dashed lines indicate sublattice ordered patterns as described in the text.}
\label{RC_ST}
\end{figure}

The above analysis is repeated for other values of $S$ at the same value of noise. The results are summarized in Fig.~\ref{RC_ST} where the solid lines illustrate the frequency of $C_e$ and $C_f$ strategies in the absence of sublattice ordering (that is, when $\rho_{1,Ce}+\rho_{1,Cf}=\rho_{2,Ce}+\rho_{2,Cf}$) meanwhile the dashed lines show the cases of $\rho_{1,Ce}+\rho_{1,Cf}>\rho_{2,Ce}+\rho_{2,Cf}$. For the latter curves the prevalence of $D_e$ and $C_f$ (or $D_f$ and $C_e$) strategies are indicated by the variation of colors. More precisely, different (green and blue) dashed lines are used to distinguish the cases of $\rho_{1,Ce}>\rho_{1,Cf}$ from the reversed relation. The readers can observe similar series of transitions in Fig.~\ref{phtr15} if $T>1$. 

For $-1 < S < 1$ the numerical analysis shows only one transition from the well-mixed state of the $C_e$ and $C_f$ strategies [$\rho_{1,Ce}=\rho_{2,Ce}=\rho_{1,Cf}=\rho_{2,Cf}$=1/2] to a sublattice ordered phase where $\rho_{1,Cf}$ and $\rho_{2,De}$ goes to 1 when $T$ is increased from 0 to 3. A third type of transition scenario can be observed in Fig.~\ref{RC_ST} in the zero noise limit if $S<-1$. In the latter case the high level of cooperation drops suddenly to a low level state at a threshold value of $T$ increasing linearly with $S$. For the latter state $D_e$ and $D_f$ strategies form a well-mixed spatial arrangement. The further increase in $T$, however, yields a continuous increase in $\rho_{1,Cf}=\rho_{2,Cf}$ and above a second threshold value of $T$ the system transforms into a sublattice ordered state where $\rho_{1,Cf}$ and $\rho_{2,De}$ approaches 1 in the large $T$ limit.

All the above mentioned continuous phase transitions consist of two consecutive phenomena when $T$ is increased. First the level of cooperation approaches 1/2 and afterward the sublattice ordered structures are built up on the analogy of anti-ferromagnetic ordering belonging to the Ising universality class \citep{stanley_71}. The effect of noise level ($K$) is also studied for several values of parameters. It is found that the width of transition regions is proportional to $K$ in both sides of the critical transition point. Consequently, the $K$ dependence of the above mentioned phases disappear in the zero noise limit and this fact allows us to determine the phase diagram with the use of a simple stability analysis as detailed in Section \ref{stabanal}.

\section{Phase diagram in the low noise limit}
\label{sec:pdlnl}

Before discussing the above results in the zero noise limit ($K \to 0$) we briefly survey some general features of this model if only two strategies are permitted. The simplest behavior can be observed when all the players are fraternal. In the latter case $C_f$ strategies prevail the whole system if $T+S<2$ while the $D_f$ and $C_f$ strategies form a chessboard like structure in the opposite case ($T+S>2$). For both structures the system achieves the maximum average payoff, that is, the fraternal players eliminate the dilemma \citep{szabo_jtb12}. 
\begin{figure}[ht]
\centerline{\epsfig{file=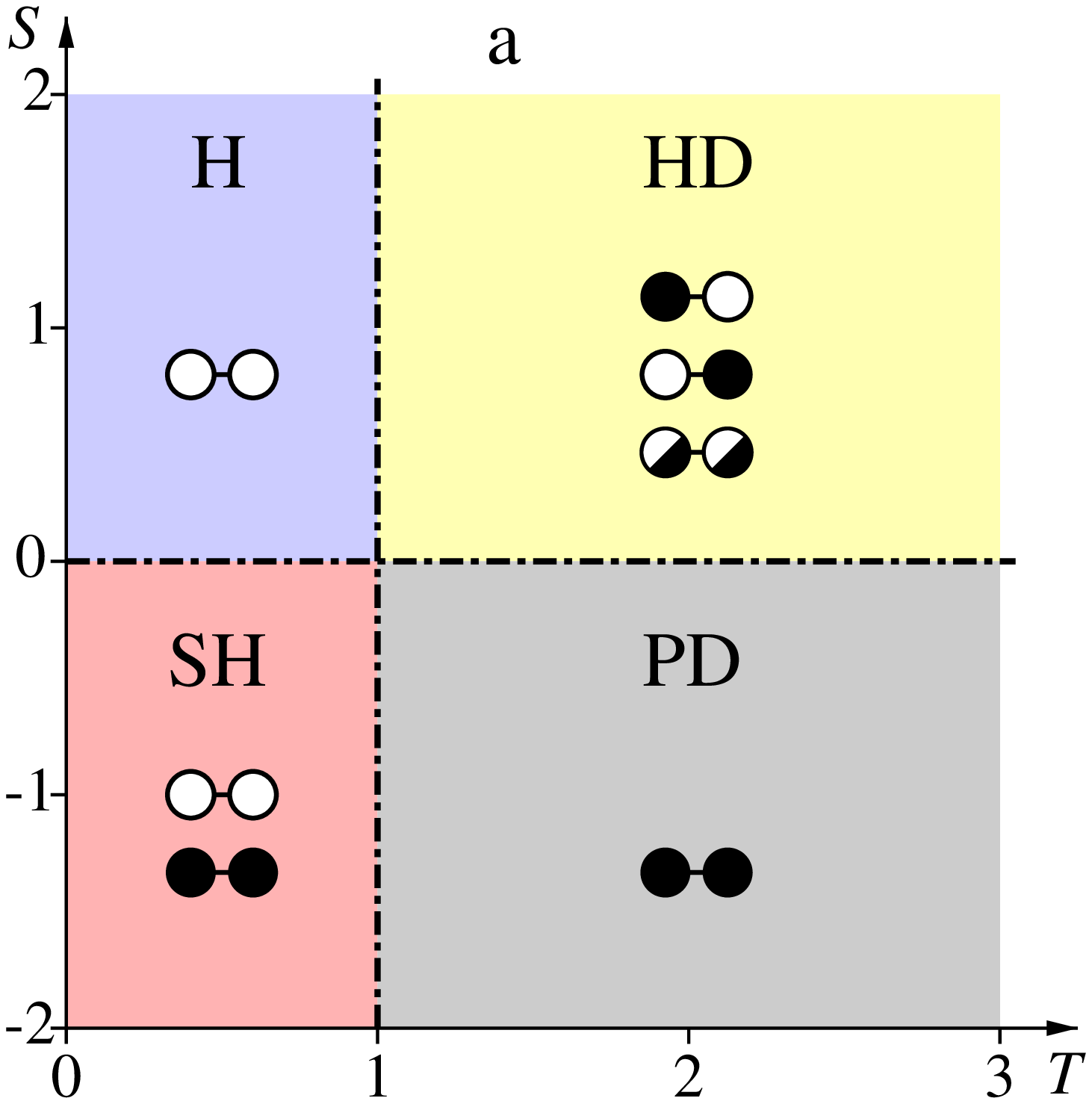,width=4.1cm}  \epsfig{file=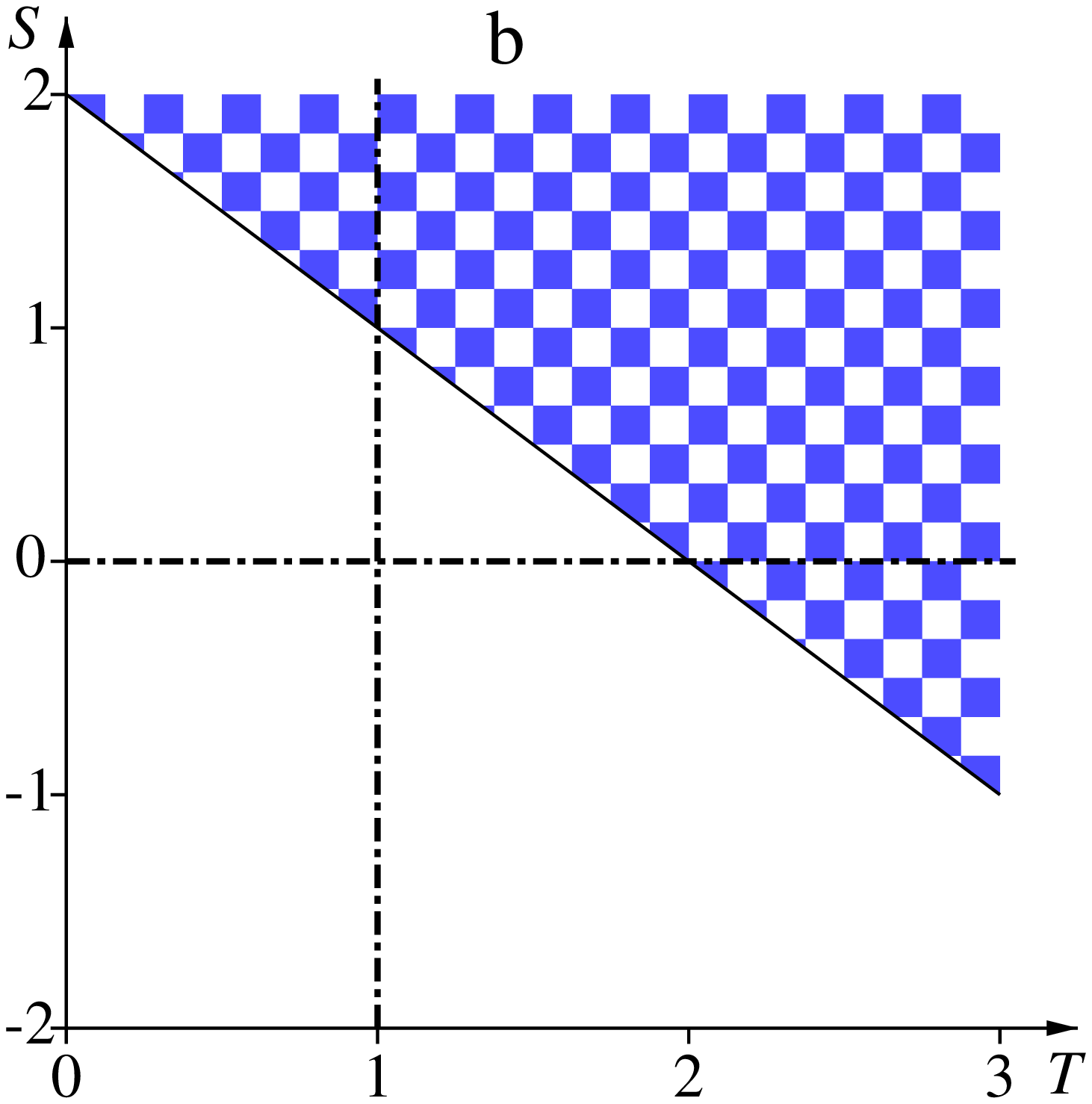,width=4.1cm}}
\vspace{0.1cm}
\centerline{\epsfig{file=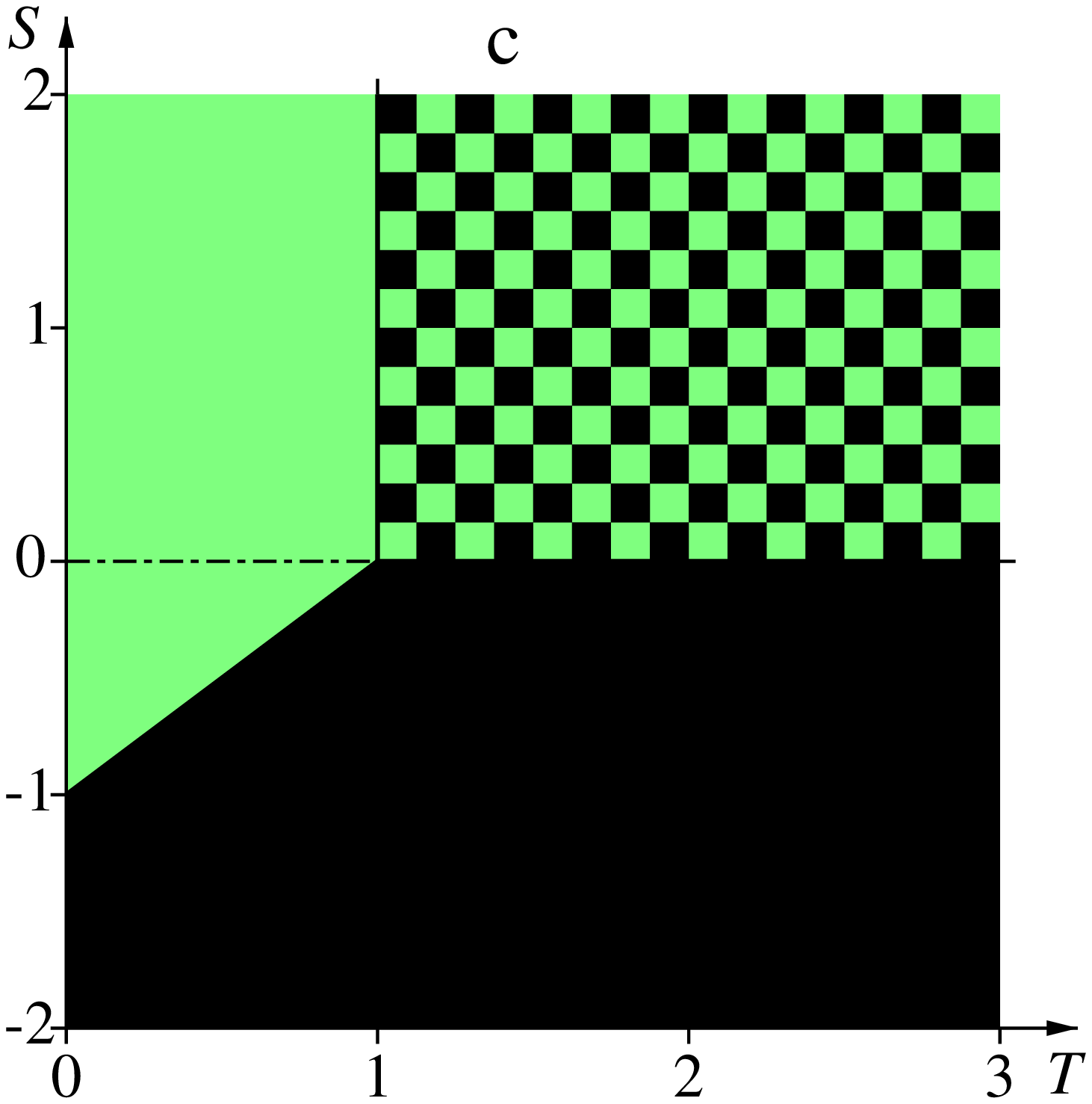,width=4.1cm}  \epsfig{file=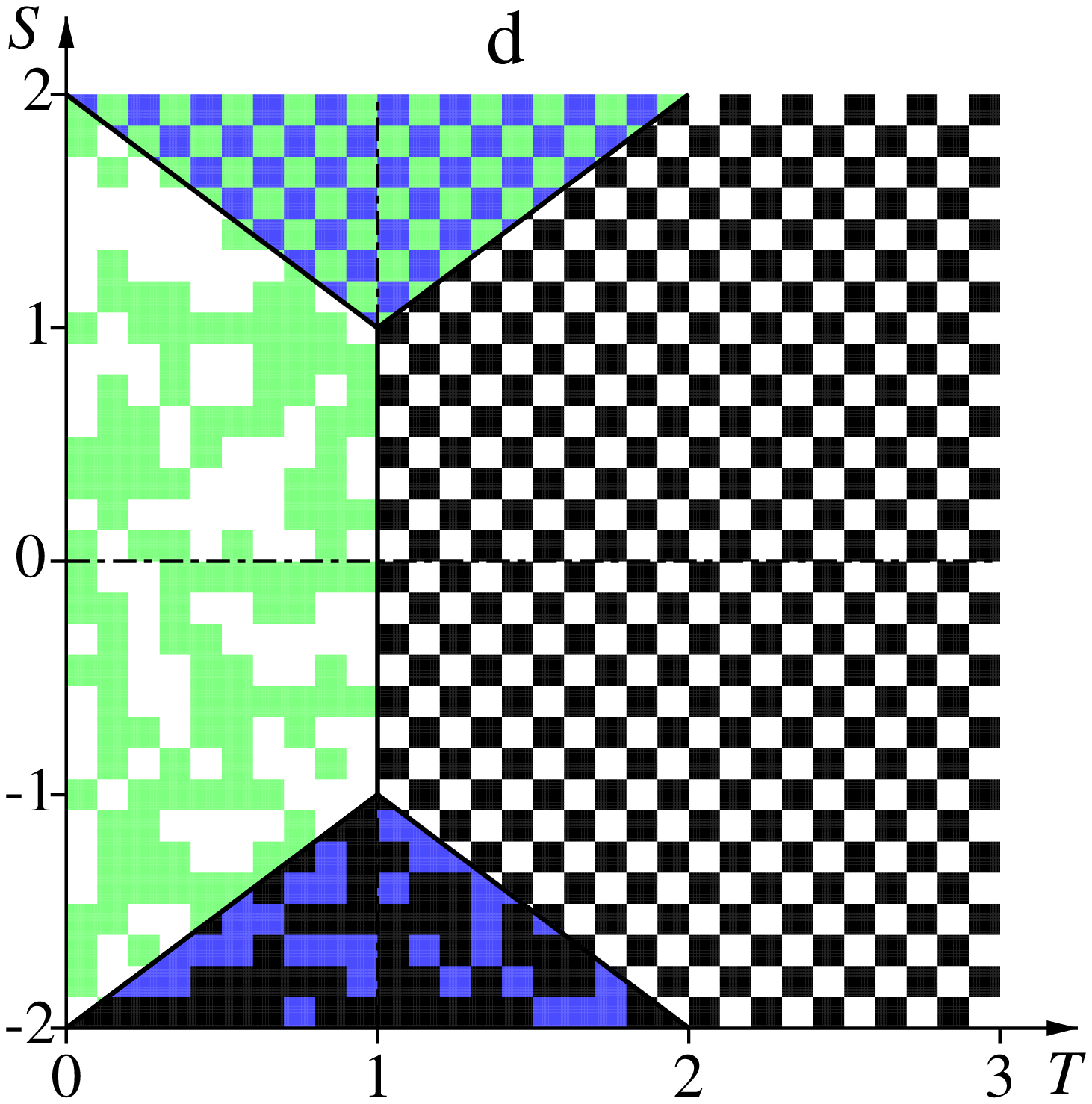,width=4.1cm}}
\caption{Solutions on $T-S$ plane under different conditions: (a) Nash equilibria for two-person one-shot games. The empty, closed, and half-filled circles indicate cooperative, defective, and mixed strategies while the regions of Harmony (H), Hawk-Dove (HD), Stag Hunt (SH) games, and Prisoner's Dilemma (PD) are separated by dashed-dotted lines. Phase diagrams in the zero noise limit if only fraternal (b), only egoist (c), and both types of players (d) are present. The colored homogeneous, random, and sublattice-ordered structures are indicated by the corresponding pattern with the same colors as in Fig.~\ref{config4}.}
\label{fig:multi}
\end{figure}

On the contrary, if all the players are egoist then the strategies $D_e$ and $C_e$ form a chessboard like (sublattice ordered) pattern ({\it i.e.}, $\rho_{1,De}=1$ and $\rho_{2,Ce}=1$) when $T>1$ and $S>0$, otherwise the cooperators ($\rho_{1,Ce}=\rho_{2,Ce}=1$ if $T<1$ and $S>T-1$) or the defectors ($\rho_{1,De}=\rho_{2,De}=1$ if $S<0$ and $S<T-1$) form a homogeneous structure \citep{szabo_jtb12}. 
The above results are summarized in Figs.~\ref{fig:multi}b-c, where the colored patterns illustrate the spatial distribution of strategies in the zero noise limit meanwhile the first plot (a) indicates the possible Nash equilibria on the four segments of the $T-S$ parameter plane for the traditional two-person one-shot games. Here it is worth mentioning that the terminology of the defective and cooperative strategies is not adequate within the region of Harmony game.

In the presence of all the four strategies the phase diagram changes drastically for the zero noise limit as indicated in Fig.~\ref{fig:multi}d where the corresponding phase boundaries are consistent with the results of the analytical stability analysis detailed in the following section. Here one can observe four two-strategy phases that are missing when only egoist or fraternal behaviors are permitted (compare the plot d with b or c). It means, at the same time, that the so-called egoist and fraternal players coexist with equal portion in all the four phases. The coexistence of $C_e$ and $C_f$ (as well as $D_e$ and $D_f$) strategies in a random time-dependent structure is a direct consequence of the equal payoff they receive within the given structures. It is more important, however, that the boundary separating the well-mixed $C_e+C_f$ and $D_e+D_f$ phases is shifted downward (compare the plots c and d in Fig.~\ref{fig:multi}) due to the presence of fraternal players. The latter positive effect of the fraternalism on the elimination of the tragedy of the commons can also be observed in the downward shift (and rotation) of the other phase boundary (for $T>1$) limiting the territory where only $D_e$ and $D_f$ strategies can survive.

In the four-strategy phase diagram (see Fig.~\ref{fig:multi}d) there is another striking result. Namely, we have found two twofold degenerated sublattice ordered structures that are different from those plotted in Figs.~\ref{fig:multi}b and c. For both structures the egoist players achieve the higher utility, that is, they exploit the fraternal individuals by choosing a suitable option. Within the upper region ($S>T$ and $T+S>2$) $D_f$ and $C_e$ strategies build up a chessboard-like structure, while in the right-hand region ($T>1$, $S<T$, and $S>-T$) the $D_e$ and $C_f$ players form a similar stable long range ordered spatial arrangement. This coexistence of the $D_e$ and $C_f$ strategies, however, can not be maintained if the sucker's payoff becomes less than a threshold value, namely, if $S<-T$.

\section{Stability analysis}
\label{stabanal}

\begin{figure}[ht]
\centerline{\epsfig{file=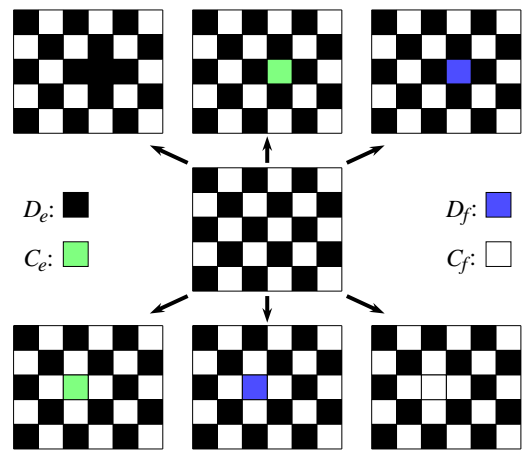,width=7.0cm}}
\caption{In the sublattice ordered arrangements of the $D_e$ and $C_f$ strategies (focal plot) six types of point defects can be created as demonstrated by the surrounding patterns.}
\label{fig5}
\end{figure}

The above mentioned analytical expressions of the phase boundaries can be derived by studying the stability of the ordered strategy arrangements against the consecutive creation of different point defects in the low noise limit. This method proved to be efficient for the investigation of a similar two-strategy system studied in a previous paper \citep{szabo_jtb12}. Now we briefly survey the extended version of this approach by considering what happens when we create different point defects in the chessboard-like arrangement of the $D_e$ and $C_f$ strategies as illustrated in Fig.~\ref{fig5}. 

First we study the case when a $C_f$ strategy is substituted for one of the $D_e$ strategy in the sublattice ordered structure of $D_e$ and $C_f$. For the present dynamical rule the given player favors this strategy change if it is beneficial for her, namely, if $U(D_e)=4T < U(C_f)=4$, where $U(D_e)$ and $U(C_f)$ denotes the utility of the given player in the initial and final constellation. The iteration of this elementary step (for any other $D_e$ players) drives the system into the homogeneous state where ${\bf s}_x=C_f \; \; \forall x$ if $T<1$, because the consecutive changes are performed in one of the sublattices and do not influence each other. Conversely, in the homogeneous state of $C_f$ the players of sublattice 1 favor the $D_e$ strategy if $T>1$ and the iteration of these elementary steps yields the chessboard like structure (see the focal plot of the Fig.~\ref{fig5}) where $\rho_{1,Cf}=\rho_{2,De}=1$. Evidently, if players in the other sublattice ($\alpha=2$) are allowed to modify their strategy then the system evolves into the so-called anti-chessboard like structure ($\rho_{2,Cf}=\rho_{1,De}=1$). If any player can modify her strategy then the iteration of the individual strategy changes yields a poly-domain structure of the two equivalent ordered patterns. In the presence of noise, however, this poly-domain structure evolves into one of the long range ordered ones through a domain growing process \citep{bray_ap94}. 

Consequently, a limited segment of the line $T=1$ can be considered as a phase boundary separating the sublattice ordered arrangement of the $D_e$ and $C_f$ strategies from the well mixed state of $C_e$ and $C_f$ players in the $T-S$ plane. The lower segment of this phase boundary is derived by studying what happens in the latter sublattice ordered strategy arrangement ($D_e+C_f$) if $D_e$ and/or $D_f$ strategies are substituted for the $C_f$s in a way described above. The similar analysis gives another phase boundary ($S=-T$, for $T>1$) separating the sublattice ordered structure of the $D_e$ and $C_f$ strategies from the region of "the tragedy of the commons" where only the defective $D_e$ and $D_f$ players are present. 

We do not wish to discuss all the possible cases we studied. It is worth mentioning, however, that a similar stability analysis of the sublattice ordered structure of the $D_f$ and $C_e$ strategies against the substitution of $C_e$ or $C_f$ for $D_f$ gives the third phase boundary ($S=2-T$ for $T<1$) mentioned above. Up to now we have discussed the derivation of three phase boundaries exhibiting a common feature. Namely, if these boundaries are crossed by increasing $T$, we can observe similar symmetry breaking. More precisely, a random structure transforms into a sublattice ordered one. For low noises these transitions exhibit similar features characterizing the Ising universality class. Deviations from this general behavior can occur in the vicinity of the tricritical points ($T=1$ and $S=1$ or $S=-1$) as well as for high levels of noise when all the four strategies are present with sufficiently large probabilities. 
 
The phase boundary $S=T$ ($T>1$) is curious. Along this line the straightforward application of the above stability analysis predicts the appearance of metastable states. For example, in the sublattice ordered arrangement of the $D_e$ and $C_f$ strategies (see the focal plot of Fig.~\ref{fig5}) the random substitution of $D_f$ for $D_e$ is favored if $S>T$ and the iteration of this process results in another sublattice ordered pattern where $\rho_{1,Cf}=\rho_{2,Df}=1$. The latter structure, however, is unstable against the strategy change from $C_f$ to $C_e$. This means that the transition from the sublattice ordered structure of the $D_e + C_f$ strategies to the $D_f + C_e$ is performed via the formation of a metastable state if $T$ decreased. The story does not ends here, because in the opposite direction (when $T$ is increased) the transformation is mediated by the formation of another metastable phase where the $C_e$ and $D_e$ strategies form a chessboard like structure. The domains of these metastable structures are recognizable in the snapshot Fig.~\ref{config4}c. 

\begin{figure}[ht]
\centerline{\epsfig{file=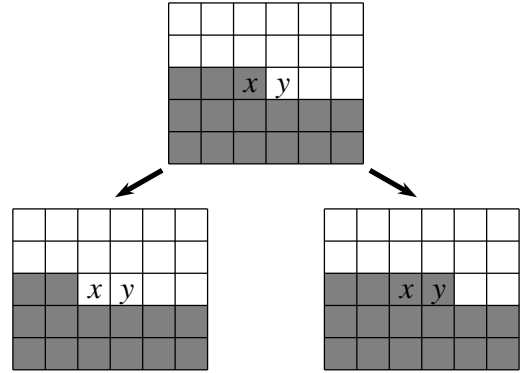,width=7.0cm}}
\caption{Two elementary steps playing determinant role in the evolution of interface separating the gray and white phases.}
\label{fig6}
\end{figure}

The determination of the phase boundary within the region of the Stag-Hunt game requires a completely different approach because here the previous stability analysis predicts two stable phases. Namely, in the well mixed structure of the defective ($D_e$ and $D_f$) strategies the present dynamical rule does not support the creation of a solitary cooperator ($C_e$ or $C_f$). Similarly, the creation of a solitary defector ($D_e$ or $D_f$) is also prohibited in the random distribution of the $C_e$ and $C_f$ strategies in the low noise limit. The situation is analogous to case described well by the ferromagnetic Ising model. In the absence of an external magnetic field the ferromagnetic system exhibits growing domains of the two stable structures. During this process the most relevant two elementary steps (see Fig.~\ref{fig6}) are executed at random and yield a slow domain growth in the absence of the external magnetic field. In the presence of the external magnetic field, however, one of these two steps is preferred and one can observe an invasion process whose direction is determined by the variation of total energy.
On the analogy of the above mentioned picture we assume that the white area in Fig.~\ref{fig6} is filled by $C_e$ and $C_f$ strategies at random and $D_e$ and $D_f$ strategies are located on the sites of the gray territory. In that case the average utility at the defective $x$ site can be given as $\langle U_d(x) \rangle = T+\sigma$ meantime the average utility at the cooperative $y$ site is $\langle U_c(y) \rangle = 2+S+\sigma$. If $\langle U_d(x) \rangle > \langle U_c(y) \rangle$ then the (gray) territory of defectors expands and finally the system evolves into the well mixed state of the $D_e$ and $D_f$ strategies. In the opposite case the cooperative ($C_e$ and $C_f$) strategies will invade the whole system. In short, this simple picture predicts a phase boundary $S=T-2$ ($T<0$) in the $S-T$ plane in agreement with a numerical results. Evidently, the present scenario requires the creation of a nucleon (with two opposite steps) along the straight line interfaces via the noisy effects. 

Interestingly, the above criterion of the selection of the winning phase coincides with the prediction of risk dominance introduced by Harsanyi and Selten \citep{harsanyi_88}. When applying the suggestion of the risk dominance for the symmetric games we assume that the players know nothing about which strategy the opponent(s) will choose. In that case the possible opponent's strategies are assumed to be chosen with equal probabilities and in the spirit of risk dominance the players favor the choice providing the higher income. Players $x$ and $y$ (in the upper plot of Fig.~\ref{fig6}) are testing the incomes under the same conditions and the favored choice determines the direction of invasion.

\section{Summary}
\label{summary}

In order to study the spatial competition between the fraternal end egoist players in the social dilemmas we have introduced a four strategy evolutionary game where the utilities are related to the two traditional parameters of the two-strategy games by revaluating the utilities with respect to their own personality. The evolution of strategy distribution is governed by a (Glauber type) myopic strategy update allowing the choice of any other strategies (involving changes in personality) with a probability dependent on the variation of the utility between the initial and final choice. This multi-agent model is analyzed by MC simulations on a square lattice for a low noise level. The resultant behavior in the final stationary state could be well reproduced by a simple stability analysis in the low noise limit.

The above investigations have justified that the present system evolves into one of the four two-strategy phases where the egoist and fraternal players coexist by forming either ordered or disordered spatial strategy distributions in the low noise limit. It is more important, however, that the presence of fraternal players has shrunk the territory of "the tragedy of the commons". The latter result can be interpreted as a consequence of the evolutionary processes favoring the coexistence of the selfish (or exploiting) players with the fraternal ones who make a sacrifice for the survival of cooperation. 

This work was supported by the Hungarian National Research Fund (grant K-101490).

\bibliographystyle{elsarticle-harv}

\end{document}